\documentclass[
superscriptaddress,
twocolumn,
 amsmath,amssymb,
 aps,
 floatfix,
prl,
]{revtex4-1}
\usepackage{graphicx,color}
\usepackage{dcolumn}
\usepackage{bm}
\usepackage{physics}

\begin{document}
\preprint{cond-mat}
\title{Fragmented Electronic Spins with Quantum Fluctuations in Organic Mott Insulators near Quantum Spin Liquid}
\author{S.\ Fujiyama}
\email[Electronic address :]{fujiyama@riken.jp}
\affiliation{RIKEN, Condensed Molecular Materials Laboratory, Wako 351-0198, Japan}
\author{R.\ Kato}
\affiliation{RIKEN, Condensed Molecular Materials Laboratory, Wako 351-0198, Japan}
\date{Sep 13, 2019}
\begin{abstract}

Magnetic structures of organic Mott insulators \textit{X}[Pd(dmit)$_{2}$]$_{2}$ (\textit{X}=Me$_{4}$P, Me$_{4}$Sb), of which electronic states are located near quantum spin liquid (\textit{X}=EtMe$_{3}$Sb), are demonstrated by $^{13}$C NMR. Antiferromagnetic spectra and nuclear relaxations show two distinct magnetic moments within each Pd(dmit)$_{2}$ molecule, which cannot be described by single band dimer-Mott model and requires intramolecular electronic correlation. This unconventional fragmentation of $S=1/2$ electron spin with strong quantum fluctuation is presumably caused by nearly degenerated intramolecular multiple orbitals, and shares a notion of quantum liquids where electronic excitations are fractionalized and $S=1/2$ spin is no longer an elementary particle.
\end{abstract} 
\pacs{75.25.-j, 75.70.Tj, 71.30.+}
\maketitle
Molecular solids exhibit remarkable quantum phenomena including $T>200$ K high-temperature superconductivity under high pressure~\cite{Drozdov2015}, switchable superconductivity by light~\cite{Suda2015}, and quantum spin liquids (QSLs)~\cite{Kanoda2011}. Of these, QSLs caused by strong quantum fluctuation are a new state of matter. This state was first theoretically proposed more than 40 years ago by considering an $S=1/2$ triangular lattice antiferromagnet, for which the geometrical frustration avoids the Neel type classical ordering~\cite{Anderson1973,Balents2010}. The strong quantum fluctuation was therefore proposed to host the exotic excitation of fractional spins~\cite{Balents2002,Han2012,Yoshitake2016}. Despite attempts to probe this hypothesis over many years, no ideal system has yet been found.
\begin{figure*}
\includegraphics*[width=14cm]{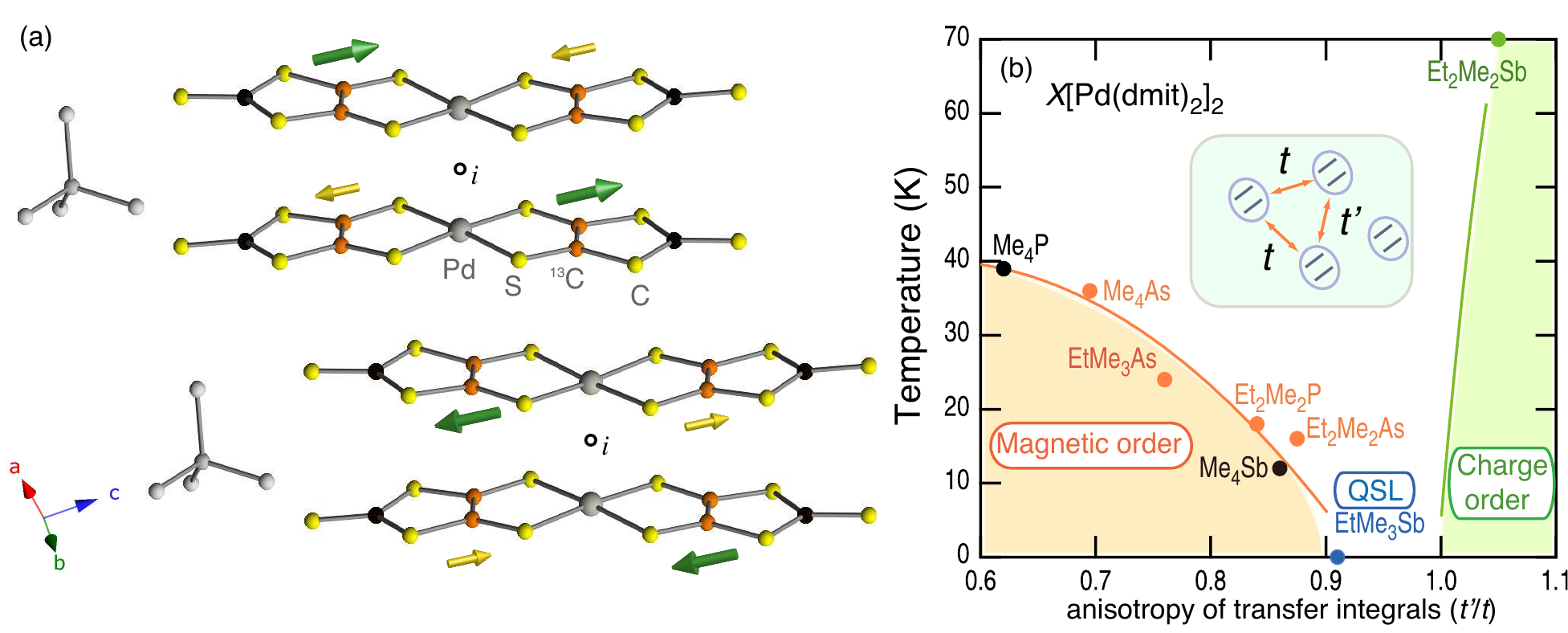}
\caption{(a) Crystal structure of \textit{X}[Pd(dmit)$_{2}$]$_{2}$ (\textit{X}=Me$_{4}$P and Me$_{4}$Sb). The dimer unit is located on the inversion center. The magnetic structure revealed by $^{13}$C NMR is demonstrated as green and yellow arrows that are fragmented $S=1/2$ spin hosted by [Pd(dmit)$_{2}$]$_{2}$ dimer. (b) Magnetic phase diagram of \textit{X}[Pd(dmit)$_{2}$]$_{2}$ as a function of the anisotropy of the transfer integrals ($t'/t$) around the quantum spin liquid (QSL) material \textit{X} = Et$_{3}$MeSb ($t'/t \sim 0.9$). The materials discussed in this article are indicated as black solid circles. In the inset, the triangle network of the interdimer transfer integrals $t$ and $t'$ between Pd(dmit)$_{2}$ dimers is shown.}
\label{fig:crystal}
\end{figure*}

In recent years, however, some possible systems in molecular solids have been proposed. Well-examined materials are $\kappa$-(BEDT-TTF)$_{2}$Cu$_{2}$(CN)$_{3}$ (BEDT-TTF = bisethylenedithio-tetrathiafulvalene) and EtMe$_{3}$Sb[Pd(dmit)$_{2}$]$_{2}$ (dmit = 1,3-dithiole-2-thione-4,5-dithiolate)~\cite{Shimizu2003,Itou2008,Kanoda2011}. The series of Mott insulators \textit{X}[Pd(dmit)$_{2}$]$_{2}$ (\textit{X}$^{+}$ = onium cation) exhibit various magnetic ground states as shown in Fig.~\ref{fig:crystal} (b)~\cite{Tamura2009}. Here, the $S=1/2$ electronic spins are localized at [Pd(dmit)$_{2}$]$_{2}$$^{-}$ dimers that are located on inversion centers, and the triangular network of interdimer transfer integrals induces spin frustration and describes the magnetic properties. \textit{X} can be chosen to tune the anisotropy of the transfer integrals ($t'/t$) of the isosceles triangles. Near the equilateral point, \textit{X} = EtMe$_{3}$Sb ($t'/t\sim 0.9$) was found to show an unconventional QSL state, in which the magnetic system avoids the classical antiferromagnetically ordered (AF) state. Here, continuous spin excitation spectra are suggested from the magnetization, thermal conductivity, and specific heat with the Wilson ratio ($R_\mathrm{w}\equiv \chi/\gamma$) $\sim1.1$ [emu J$^{-1}$ K$^{2}$]), which is consistent with the concept of the long-range RVB (resonating valence bond) hypothesis~\cite{SYamashita2008,MYamashita2010,SYamashita2011,Watanabe2012}. Despite the fermionic character of spins (spinons) suggested from these macroscopic measurements, discrepancies remain between these measurements and the nuclear spin-lattice relaxation rate, $1/T_{1}$, obtained by $^{13}$C NMR~\cite{Itou2008}. The reported $1/T_{1}$, which is proportional to $T^{2}$, suggests a small Fermi surface as has been observed for anisotropic superconductors. $1/T_{1}$ also has a kink at $T\sim 1$ K, the reason for which is unknown, however, a similar anomaly has not been observed in macroscopic measurements.

Distinct from the electronic spins localized at the lattice points in inorganic compounds, those in molecular solids are more widely spatially distributed by a factor of $\sim 15$ (the average space per spin is 840 \AA$^{3}$ for \textit{X}[Pd(dmit)$_{2}$]$_{2}$ and 60 \AA$^{3}$ for ZnCu$_{3}$(OH)$_{6}$Cl$_{2}$). In molecular conductors, the tight-binding approximation, which treats the $\pi$-electrons on a molecular dimer as simply $S=1/2$ electronic spins, has been applied to analyze the physical properties. However, it is possible for the intramolecular electronic correlation to impact the physical properties of the material. In this context, it is still an open question whether the QSLs of inorganic and organic materials are subsumed under a universality class, or are characteristic to individuals.

In this Letter, we show evidence that the intramolecular correlation plays a vital role in the magnetism of Pd(dmit)$_{2}$ salts near the quantum spin liquid state. The NMR spectra and nuclear spin relaxations of the salts that undergo antiferromagnetic ordering require intramolecular fragmentation of $S=1/2$ electron spin with strong quantum fluctuations. The determined magnetic structure cannot be described by the dimer-Mott models, and presumably caused by nearly degenerated intramolecular multiple orbitals. We discuss the internal degree of freedom behind electron spins can be a source to stabilize QSL state in this salts and resolve the discrepancy between macroscopic measurements and $1/T_{1}$ for the QSL (\textit{X}=EtMe$_{3}$Sb). 

Single crystals of \textit{X}[Pd(dmit)$_{2}$]$_{2}$ (\textit{X} = Me$_{4}$P and Me$_{4}$Sb) were synthesized by air oxidation of \textit{X}$_{2}$[Pd(dmit)$_{2}$] in an acetone solution containing acetic acid~\cite{Kato2012}. We labelled one of the carbon atoms of the C=C bond in the dmit ligand (shown in red in Fig.~\ref{fig:crystal} (a)) by $^{13}$C. $^{13}$C NMR spectra were obtained by the Fourier transforms (FTs) of the spin echo for the paramagnetic state and by convolution of the  FT spectra obtained by shifting the resonating frequency using randomly oriented crystals. Nuclear relaxation measurements were performed using a single crystal by applying external magnetic fields along the $c$-axis, and $1/T_{1}$ shows a marked peak at $T_\mathrm{N} =$ 40 K (12 K) for \textit{X} = Me$_{4}$P (Me$_{4}$Sb) reflecting magnetic ordering.

\begin{figure*}
\includegraphics*[width=17cm]{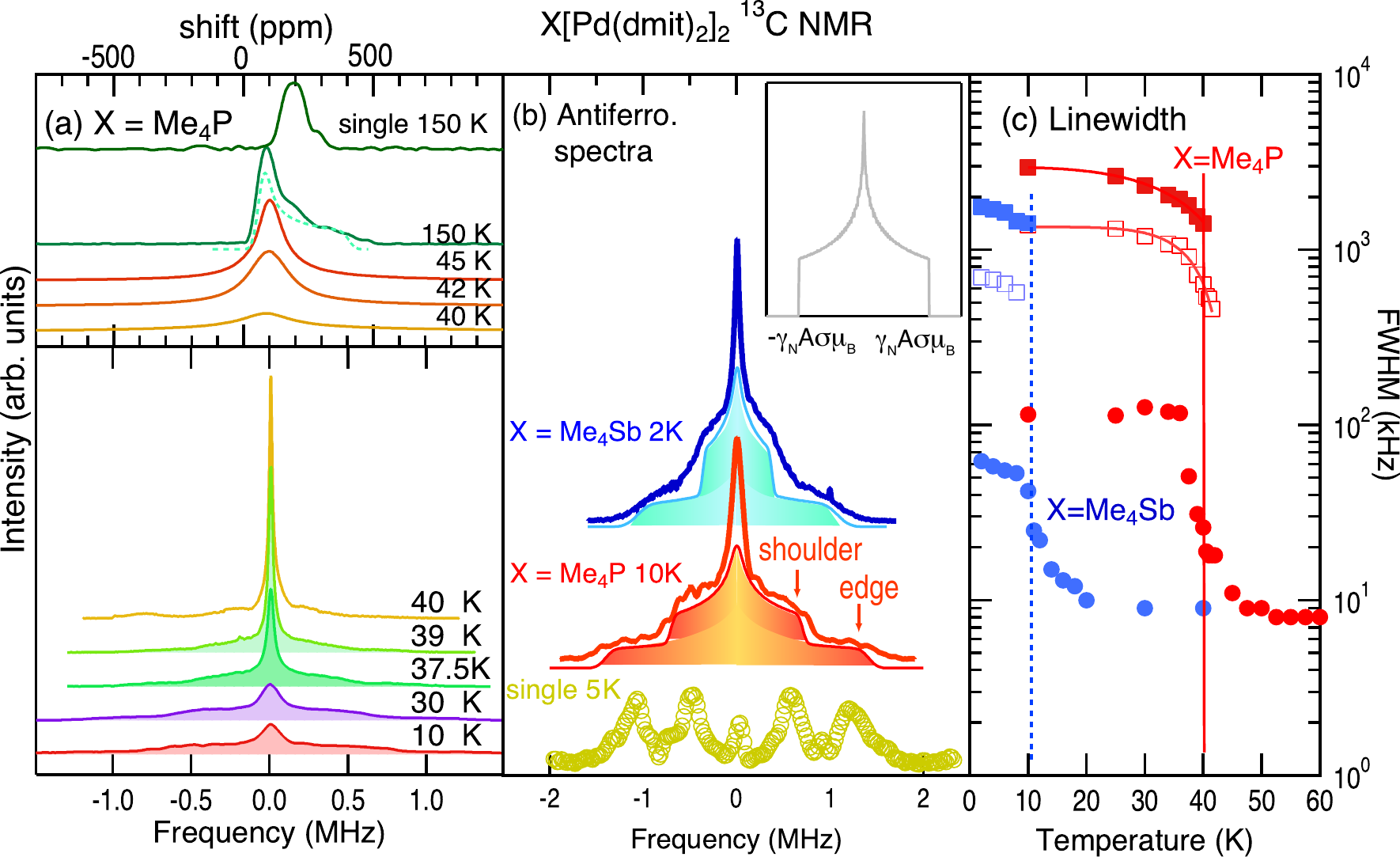}
\caption{(a) Temperature-dependent powder-averaged $^{13}$C NMR spectra of Me$_{4}$P[Pd(dmit)$_{2}$]$_{2}$ and a single crystal spectrum at 150 K. The spectra at 40 K are plotted in both panels for the paramagnetic and ordered states. The Larmor frequency and the resonance of TMS (tetramethylsilane) are set to 89.5 MHz. A calculated spectrum using the values of reported susceptibility at 150 K ($\chi=5 \times 10^{-4}$ [emu/mol])~\cite{Tamura2002} and hyperfine coupling constant ($A=2.5$ [kOe/$\mu_{B}$]) is plotted as dashed line. (b) Antiferromagnetic resonance spectra for \textit{X} = Me$_{4}$P and Me$_{4}$Sb at temperatures where the magnetic moments are nearly saturated. Solid lines show calculated spectra assuming two distinct fractional magnetic moments, $\sigma=0.53$ and 0.28 $\mu_\mathrm{B}$ for \textit{X}=Me$_{4}$P and 0.31 and 0.12 $\mu_\mathrm{B}$ for Me$_{4}$Sb. The inset shows a spherically averaged antiferromagnetic spectrum for one magnitude of the magnetic moment. The open circles show spectra using a single crystal at 5 K, in which peaks correspond the edges and shoulders in the powder averaged spectrum. A small amount of residual signal of grease is visible at $\sim 0$ MHz. (c) Full width at half maximum of the peak at the central frequency (89.5 MHz) (solid circles) and the frequency differences between the edges and shoulders of the spectra (solid and open squares).
}
\label{fig:spectra}
\end{figure*}

Temperature-dependent NMR spectra are displayed in Fig.~\ref{fig:spectra} (a) for the \textit{X}=Me$_{4}$P salt. Asymmetric spectra at 150 K are well reproduced assuming hyperfine coupling constants with axial symmetry. We estimate the onsite hyperfine coupling constants of four independent $^{13}$C sites with nearly the same values ranging from $2.2$ to $2.6$ kOe/$\mu_{B}$ by DFT (density functional theory) calculations of the spin density at each site. The values are similar to those of the central carbon sites of $\kappa$-BEDT-TTF compounds~\cite{Kawamoto1995}. The spectra using a single crystal at 150 K is almost symmetric and sharp of which full width at the half maximum is 100 ppm, which supports nearly the same values of hyperfine coupling constants as shown in Fig.~\ref{fig:spectra} (a). Below about 60 K, the asymmetry and structures are smeared out by enhanced antiferromagnetic fluctuation. Significant broadening of the spectra is observed below $T_\mathrm{N}$, and we obtained the antiferromagnetic resonance spectra shown in Fig.~\ref{fig:spectra} (b) as the ground state for \textit{X} = Me$_{4}$Sb and Me$_{4}$P. The powder averaged spectra have an unconventional structure composed of shoulders and edges, ruling out conventional collinear antiferromagnetism.

The NMR frequency in the ordered state is shifted by the sum of the onsite hyperfine fields at the $^{13}$C ($\Delta \omega_\mathrm{on}$) and dipolar fields by offsite $\pi$ electrons ($\Delta \omega_\mathrm{dip}$) as follows:
\begin{eqnarray}
\Delta\omega&=&\Delta\omega_\mathrm{on}+\Delta\omega_\mathrm{dip}\\ \nonumber
&=&\sum_{j} \gamma_{N}A\mu_{j} \qty[ \qty( 3\hat{\mu}_{j}\cdot \hat{e}_{\pi})\qty( \hat{e}_{\pi}\cdot \hat{H}_{0})-\qty( \hat{\mu}_{j} \cdot \hat{H}_{0})]\\ \nonumber
&+&\sum_{j}\gamma_{N}\frac{\mu_{j}}{r_{ij}^{3}}\qty[ \qty( 3\hat{\mu}_{j} \cdot \hat{r}_{ij})\qty(\hat{r}_{ij}\cdot \hat{H}_{0})-\qty( \hat{\mu}_{j}\cdot \hat{H}_{0})].
\end{eqnarray}
Here, $\gamma_{N}$ is the gyromagnetic ratio of $^{13}$C, $A$ is the hyperfine coupling constant, $\displaystyle  \mu_{j}=\sigma_{j}\mu_{B} $ is the magnetic moment at site $j$, $\sigma_{j}$ is the fractional spin density, $\mu_\mathrm{B}$ is the Bohr magneton, $\hat{e}_{\pi}$ is the unit vector of $\pi$ electrons normal to the molecular plane, and $\hat{H}_{0}$ is the unit vector of the external fields.  $\Delta\omega_\mathrm{dip}$ is found to be limited to as small as 1/10 of $\Delta\omega_\mathrm{on}$ and is neglected in the following fitting of the spectra.

We apply a spherical average to the equation and plot the simulated spectrum for one magnitude of the magnetic moment $\mu=\sigma \mu_{B}$ in the inset of Fig.~\ref{fig:spectra} (b), which shows a peak at the center and cut-off edges at $\pm \gamma_{N}\sigma A\mu_{B}$. To reproduce the observed spectra, we require two distinct $\mu_{j}$ with nearly the same intensity. We fitted the spectra and obtained 0.53 and 0.28 $\mu_\mathrm{B}$ for \textit{X}=Me$_{4}$P and 0.31 and 0.12 $\mu_\mathrm{B}$ for Me$_{4}$Sb. The absence of spontaneous magnetization according to macroscopic susceptibility measurements rules out ferrimagnetic ground state. The NMR spectra show antiferromagnetic structures with two contrasting magnetic moments, where electronic spins have to be fragmented regardless of the mechanism involved.

One question raised by the antiferromagnetic spectra is whether the electron fragmentation has been set even in the paramagnetic state or limited to below $T_\mathrm{N}$. We plot in Fig.~\ref{fig:spectra} (c) the frequency differences between the shoulders and edges on both sides of the spectra in Fig.~\ref{fig:spectra} (b). The evolution of the frequency differences slightly below $T_\mathrm{N}$ is so rapid that the linewidth does not follow a $\left( T_\mathrm{N}-T\right)^{n}$ relation. We conclude that the magnetic orders have a first-order character, which suggests the existence of other channels than magnetic correlation.

\begin{figure*}[htb]
\includegraphics[width=8.5cm]{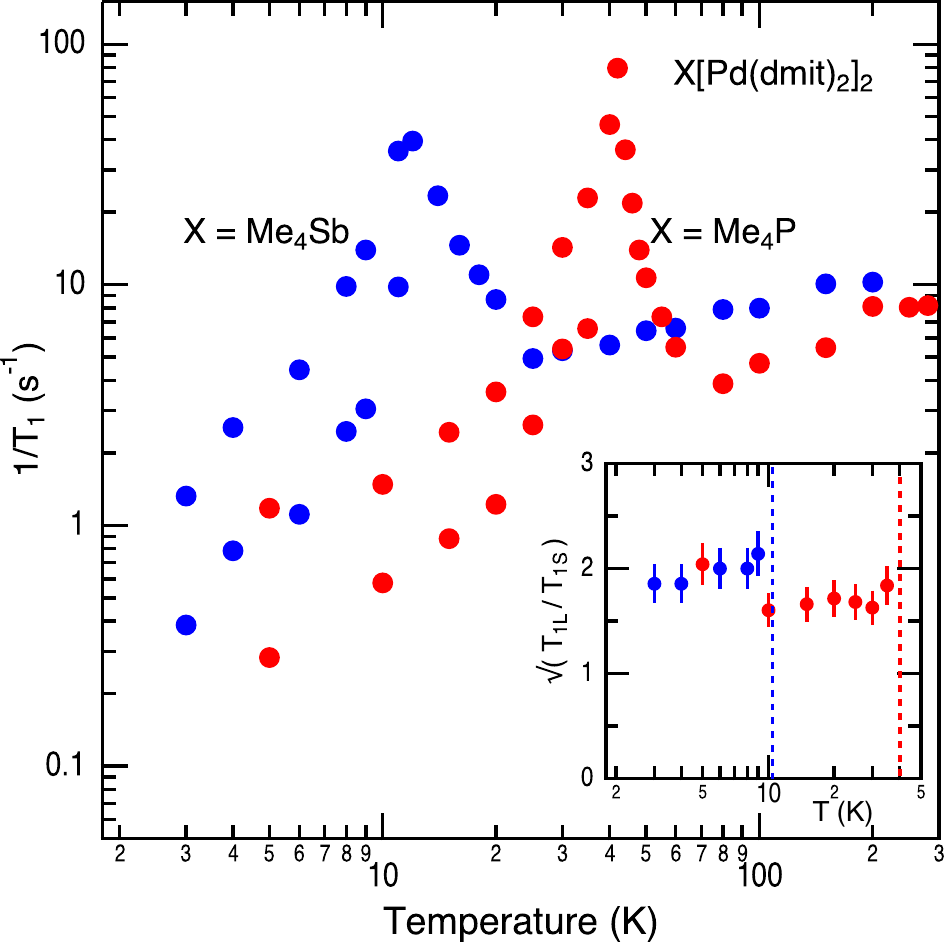}
\caption{Nuclear spin-lattice relaxations ($1/T_\mathrm{1}$) at the $^{13}$C sites using a single crystal. The divergence of $1/T_\mathrm{1}$ shows an antiferromagnetic transition at $T_\mathrm{N}=40$ K (\textit{X}=Me$_{4}$P) and $T_\mathrm{N}=12$ K (\textit{X}=Me$_{4}$Sb). In the inset, the square root of the nuclear spin-lattice relaxation rates measured at the edge ($1/T_{S}$) and shoulder ($1/T_{L}$) frequencies below $T_\mathrm{N}$ (dashed lines).}
\label{fig:T1}
\end{figure*}

We plot $1/T_{1}$ for the paramagnetic state and those measured at the edge and shoulder frequencies for the antiferromagnetic state using a single crystal in Fig.~\ref{fig:T1}. The nuclear magnetization curves can be fitted by a single exponential formula. The obtained $1/T_{1}$ measured at the edge frequency is three times as large as that at the shoulder frequency for both salts, which remains constant below $T_\mathrm{N}$.

$1/T_{1}$ is proportional to the dynamical susceptibility, $\chi''(\vb{q},\omega)$ expressed by
\begin{equation}
1/T_{1}=\gamma_{N}^{2}T\sum_{\vb{q}} A_{\perp}^{2}(\vb{q})\chi_{\perp}''(\vb{q},\omega_{0})/\omega_{0},
\end{equation}
where $\omega_{0}$ is the Larmor frequency, $A_{\perp}(\vb{q})$ and $\chi''_{\perp}(\vb{q},\omega_{0})$ are the hyperfine coupling constant and the imaginary part of the dynamical susceptibility perpendicular to the external fields, respectively.

The nearly temperature-independent $1/T_{1}$ for $T>200$ K is consistent with a Mott insulator with antiferromagnetic correlation. $1/T_{1}$ increases approaching $T_\mathrm{N}$, which is due to the slowing down of electronic spins, and has peaks at 40 K (for \textit{X}=Me$_{4}$P) and 12 K (Me$_{4}$Sb) corresponding to the phase transitions. Combined with the marked broadening of the spectra shown in Fig.~\ref{fig:spectra} (c), it is concluded that the antiferromagnetic transition has a weak first-order character. 

The distinct spin densities indicated by the antiferromagnetic resonance spectra in Fig.~\ref{fig:spectra} (b) are also supported by the relaxation measurements. The square root of the ratio $\displaystyle \sqrt{T_\mathrm{1L}/T_\mathrm{1S}}$, which is expected to be proportional to the ratio of $A$, remains $\sim 2$ below $T_\mathrm{N}$ (Fig.~\ref{fig:T1} (b)). This observation again evidences the unconventional fragmentation of electrons with a factor of 2 in the magnetically ordered state. 

Recent experiments on QSL molecular materials involving the measurements of the dielectric constant, optical conductivity, vibrational spectroscopy, and nuclear quadrupole resonance have revealed low-energy charge fluctuations~\cite{AbdelJawad2010,AbdelJawad2013,KItoh2013,Yamamoto2017,Fujiyama2018}. The charge fluctuations, whose excitation spectra are gapless, can cause softening of a spin wave and prevent classical magnetic ordering. Theoretical studies on a dimer Mott insulator with a triangular lattice also revealed ferroelectric polarization, and the exclusive relationship between the ferroelectricity and Mott antiferromagnetic state was discussed~\cite{Naka2010,Watanabe2017}.

Although this scenario appears to be consistent with the distinct spin densities in the present observations, the theoretically proposed polarization composed of charge-rich and -poor Pd(dmit)$_{2}$ molecules loses its inversion symmetry between dimerized molecules. X-ray crystal structure analysis shows no symmetry-breaking phase transition across $T_\mathrm{N}$ and the inversion symmetry remains conserved~\cite{Ueda2018,Rouzire1999}. The intramolecular distinction of the magnetic moments is required to satisfy the space group, and the magnetic structure is uniquely determined as shown in Fig.~\ref{fig:crystal} (a). Here, the magnitudes and phases of intradimer two green (yellow) arrows are constrained with an antiparallel relation to the two green (yellow) moments in the adjacent dimer. The total magnetization of the green (yellow) arrows is zero in a unit cell, which is consistent with Mott antiferromagnetism. We consider that the constraint between the green and the yellow arrows does not originate from magnetic Hamiltonians but is determined by other sources such as intramolecular Coulomb interaction.

It is to note that the determined magnetic structure requires fragmentation of the $S=1/2$ electronic spin associated with strong quantum fluctuations near the QSL state. This is considerably different from the broadly accepted dimer Mott approach to the magnetism of molecular solids, in which a simplified electronic wave function on molecules is taken the HOMO (highest occupied molecular orbital) or LUMO (lowest unoccupied molecular orbital) to form a single band.

\begin{figure}[htb]
\includegraphics*[width=6.5cm]{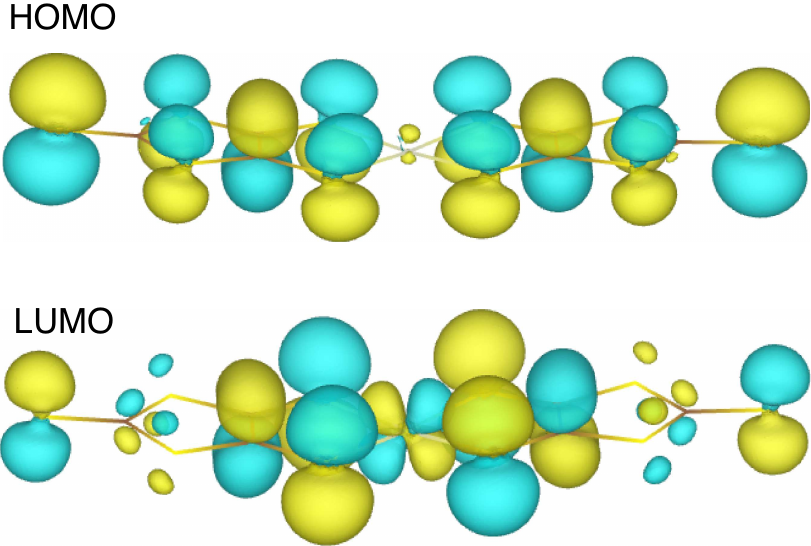}
\caption{$\pi$ molecular orbitals in the Pd(dmit)$_{2}$ monomer at the HOMO and LUMO. The ungerade symmetry of the HOMO rules out the contribution of the $4d$ electrons of the Pd ion when the molecule is isolated. The LUMO, which has gerade symmetry, exhibits $\pi$-$d$ hybridization.}
\label{fig:HOMO}
\end{figure}

One prominent characteristic of the material studied is that the constituent molecule Pd(dmit)$_{2}$ contains a central transition metal Pd. The \textit{d}-electrons of Pd seldom contribute to the HOMO because of the ungerade character of the MO as shown in Fig.~\ref{fig:HOMO}~\cite{Canadell1990}. The absence of the $\pi$-$d$ bond makes the HOMO unstable and raises its energy level and, as a result, the energy gap between the HOMO and LUMO becomes as small as $\sim 0.8$ eV, which is one-third of that of other organic molecules such as BEDT-TTF and TMTTF (TMTTF = tetramethyl-tetrathiafulvalene) salts~\cite{Tajima1991}. The strong dimerization of Pd(dmit)$_{2}$ and the lost mirror symmetry of C$_{2v}$ with respect to the left and right dmit ligands result in significant HOMO-LUMO hybridization, which may be uneven between two dmit ligands across the central Pd. The contrasting charge densities of ligands can also be evaluated by this multiorbital effect~\cite{Miyazaki1999,Tsumuraya2013}.

This intramolecular symmetry breaking is expected to be quantified by \textit{ab initio} calculation. However, a partial DOS calculation for this material gives only approximately $\displaystyle (0.47 \mu_\mathrm{B}/0.43 \mu_\mathrm{B} -1) =9$\% asymmetry in the spin densities of each dmit ligand. This small asymmetry indeed reproduces the merged spectra in the paramagnetic state, as shown in Fig.~\ref{fig:spectra} (a), and the reported NMR spectra in the antiferromagnetic state, in which outer carbon site is labelled by $^{13}$C with a very small hyperfine coupling constant~\cite{Otsuka2014}. The distinct magnetic moments with a factor of $\sim$ 2 require other sources than those taken into account by the band calculation. A plausible source is intramolecular Coulomb repulsion between the ligands, which induces an intramolecular antiferromagnetic exchange interaction.

A recent RPA (random phase approximation) calculation of the Pd(dmit)$_{2}$ salts assuming two independent fragmented molecular orbitals predicted a significant difference in the magnitudes of the magnetically ordered moments on each dmit ligand~\cite{Seo2015}. Here the ratio of the ordered moments was estimated to be as large as three, although the ratio may have been overestimated depending on the onsite Coulomb repulsion $U$. Our observations of distinct ordered moments with a ratio of up to 2.5 agree with the calculation, and the intramolecular electronic correlation is concluded to affect the magnetic state in this material near the QSL state.

We discuss the connection between the fragmented molecular orbital treatment of the electronic wave function and the accepted dimer Mott approach that describes the electronic phase diagram including the QSL as shown in Fig.~\ref{fig:crystal} (b). When we regard the electronic correlation as a triangular lattice whose lattice points are composed of a dimer of two Pd(dmit)$_{2}$ molecules, the ordered moment per dimer at each site ($\mu_{d}\equiv \abs{\mu_{\uparrow}}-\abs{\mu_{\downarrow}}$) is evaluated as $\mu_{d}=0.53-0.25=0.28$ $\mu_\mathrm{B}$ for Me$_{4}$P ($T_\mathrm{N}=40$ K) and $\mu_{d}=0.31-0.12=0.19$ $\mu_\mathrm{B}$ for Me$_{4}$Sb ($T_\mathrm{N}=12$ K) because the intramolecular magnetic correlation is considered to be antiferromagnetic. The greatly reduced value of the ordered moment of the $S=1/2$ dimer is evidence of enhanced quantum spin fluctuation approaching the QSL state at $t'/t\sim 1$. We consider that the phase diagram parameterized by $t'/t$ in Fig.~\ref{fig:crystal} (b) is still applicable even though the fragmented molecular orbitals are stabilized by the intramolecular electronic correlation.

The overall justification for the macroscopic phase diagram again reproduced by the independent magnetic moment for each ligand provides a key to resolving the contradiction between the macroscopic measurements and $1/T_{1}$ for the QSL in EtMe$_{3}$Sb[Pd(dmit)$_{2}$]$_{2}$. It is natural that the intramolecular correlation is smeared out by macroscopic susceptibility or heat measurements, while NMR is coupled to microscopic electronic states. The reported kink of $1/T_{1}$ at $T\sim 1$ K may be closely related to the instability of the intramolecular fragmentation of the electronic spins.

Recent theoretical works share a notion that the QSL can provide fractionalized spin excitations (spinons) caused by strong quantum fluctuation~\cite{Balents2002,Yoshitake2016}. The experimental observation of continuum spin excitation by neutron scattering has been argued to correspond to the fractionalization of electrons~\cite{Han2012}. Here for the molecular-based QSL, unlike in the case of inorganic magnets, the electronic spin possesses freedom of fragmentation by the multiorbital effect, which causes damping of the magnon and the resultant nonlocal character of spin excitations. The uniqueness of molecular QSL materials evincing a nonlocal order parameter of $R_\mathrm{W} \sim 1$ might be related to the local symmetry breaking in the [Pd(dmit)$_{2}$]$_{2}$ dimer with $S=1/2$. This intramolecular electronic correlation, which has been overlooked in the dimer Mott approach, at least acts as a supporting source to stabilize the QSL.


To conclude, the antiferromagnetic resonance spectra and $1/T_{1}$ for \textit{X}[Pd(dmit)$_{2}$]$_{2}$ (\textit{X}=Me$_{4}$P, Me$_{4}$Sb) show the fragmentation of $S=1/2$ electrons across $T_\mathrm{N}$. Distinct magnetic moments with a ratio of up to 2.5 have not yet been realized by the electronic correlation of a single band model, suggesting an indispensable role of the intramolecular degree of freedom. This is a unique characteristic of the Pd(dmit)$_{2}$ molecule, which contains central transition metal that reduces the HOMO-LUMO energy gap. 

This internal degree of freedom behind electrons raises a problem regarding electronic correlations in molecular solids. The magnetic phase diagram of competing orders shown in Fig.~\ref{fig:crystal} (b) must be revisited by taking the instability against spin fragmentation into account, particularly at the AF-QSL phase boundary. The weak first-order character of the boundary may extend the nonmagnetic ground states near the equilateral point of the electronic correlation, $t'/t=1$, and we consider that no specific quantum critical point can be pinned down and that a QSL is realized as a phase in the $T$-$(t'/t)$ diagram in molecular solids.

The intramolecular degree of freedom, which can play an important role in determining physical properties, may be a supporting source to stabilize the gapless QSL state. We also claim that this new degree of freedom will extend the functionality of molecular solids.

\begin{acknowledgments}
We are grateful to K.~Ueda, H.~Seo, H.~Watanabe, T.~Tsumuraya, and K.~Hiraki for helpful discussions. This work was supported by Grants-in-Aid for Scientific Research (26400378, 16H06346) from JSPS. The computations of the hyperfine coupling constants and dipolar fields were performed on the RIKEN supercomputer system.
\end{acknowledgments}
\end{document}